\documentclass[aps,prl,twocolumn,groupedaddress,showpacs]{revtex4}
\usepackage{bm}
\usepackage{epsf}
\usepackage{amssymb}
\usepackage{amsmath}
\usepackage{graphicx}
\usepackage{rotating}
\usepackage{epsfig}
\usepackage{psfrag}
\usepackage{amsmath}
\usepackage{hyperref}

\hypersetup{
    bookmarks=true,         
    unicode=false,          
    pdftoolbar=true,        
    pdfmenubar=true,        
    pdffitwindow=true,      
    pdftitle={My title},    
    pdfauthor={Author},     
    pdfsubject={Subject},   
    pdfcreator={Creator},   
    pdfproducer={Producer}, 
    pdfkeywords={keywords}, 
    pdfnewwindow=true,      
    colorlinks=true,       
    linkcolor=red,          
    citecolor=blue,        
    filecolor=magenta,      
    urlcolor=blue           
}

\DeclareMathAlphabet{\bi}{OML}{cmm}{b}{it}

\begin{document}
\def\ea{\textit{et al.}}
\def\bj{\bm{j}}
\def\Im{\mathrm {Im}\;}
\def\Re{\mathrm {Re}\;}
\def\beq{\begin{equation}}
\def\eeq{\end{equation}}
\def\Tr{{\mathrm{Tr}}}
\def\sx{\hat{\sigma}_x}
\def\sy{\hat{\sigma}_y}
\def\sz{\hat{\sigma}_z}
\def\bq{\mathbf{q}}
\def\oc{\omega^{*}}
\def\Hss{\hat{H}_{\mathrm{s}}}

\title{Quantum bounds on heat transport through nanojunctions}
\author{Edward~Taylor}
\affiliation{Chemical Physics Theory Group, Department of Chemistry, University of Toronto,
80 Saint George St. Toronto, Ontario, Canada M5S 3H6}
\author{Dvira~Segal}
\affiliation{Chemical Physics Theory Group, Department of Chemistry, University of Toronto,
80 Saint George St. Toronto, Ontario, Canada M5S 3H6}

\date{\today}

\begin{abstract}
We derive rigorous quantum mechanical bounds for the heat current through a nanojunction connecting two thermal baths at different temperatures. Based on exact sum rules, these bounds compliment the well-known quantum of thermal conductance $\kappa_Q \equiv \pi k^2_B T/6\hbar$, which provides a bound for low-temperature heat transport in all systems, but is saturated only for noninteracting transport. In contrast, our bounds are saturated at high temperatures---but still in the quantum regime---, even when interactions are very strong. We evaluate these bounds for harmonic and strongly anharmonic junction models and compare with numerical approaches.  
\end{abstract}
\maketitle

Does quantum mechanics place bounds on the rate of equilibration in a many-body system? When the initial state of the system describes an inhomogeneity in some quantity (e.g., temperature, chemical potential, velocity), this question can be cast in terms of bounds on currents and, in the linear response regime, their corresponding transport coefficients (thermal and charge conductivities, viscosity). Heuristic bounds on such coefficients based on Heisenberg's uncertainty principle have been conjectured in a wide array of systems, ranging from the quark-gluon plasma~\cite{Danielewicz85,Schaefer09} and ultra-cold atomic gases~\cite{Cao11,Bardon14}, to incoherent ``bad metals''~\cite{Hussey04,Hartnoll15}. The considerable utility of such bounds is that they can give physical insight into the nature of transport in regimes where interactions are strong and perturbative calculations of transport break down~\cite{Schaefer09,Cao11,Bardon14,Hussey04,Hartnoll15,Bruin13}. Unfortunately, as far as we know, none of these bounds have yet been made rigorous~\cite{Jung07}.  

\begin{center}
\begin{figure}
\includegraphics[width=0.7\linewidth]{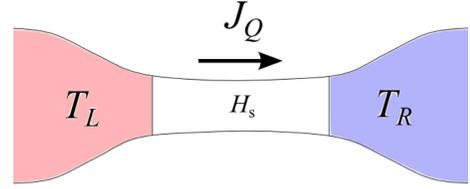}
\caption{Schematic figure of a nanojunction comprising a central subsystem described by the hamiltonian $\Hss$, connecting two thermal baths at temperatures $T_L$, $T_R$. A heat current $J_Q$ arises when $T_L\neq T_R$.}\label{fig1}
\end{figure}
\end{center}

Bounds on heat transport are of special value, connecting to foundational questions of information and entropy flow~\cite{Pendry83,Blencowe00}. At the same time, there is a pressing need to understand the limits of nanoscale devices such as semiconductor nanowires~\cite{Kim1}, carbon nanotubes~\cite{Kim2}, silicon membranes~\cite{Nelson}, and molecular chains~\cite{Segalman,Dlott,Gotsmann}, to manipulate and transport heat~\cite{Kohler05,Pop10,Dubi11,Li12,Cahill14}, an enterprise for which interactions may prove crucial~\cite{Murphy08}. For systems in which a heat current arises in a subsystem connecting two thermal baths at temperatures $T_L$ and $T_R$ (see Fig.~\ref{fig1}) with the current described by the Landauer-type expression
\beq J_Q = \frac{1}{2\pi}\int^{\infty}_{-\infty} d\omega \hbar \omega {\cal{T}}(\omega,T_L,T_R)[n_L(\omega)-n_R(\omega)],\label{Landauer}\eeq
it is well known that the heat current is bounded by~\cite{Pendry83,Sivan86,Butcher90,Blencowe00}:
\beq J_Q \leq \frac{\pi k^2_B}{12\hbar}(T^2_L-T^2_R).\label{kappaQ}\eeq
$n_L$ and $n_R$ are the Bose (or Fermi) thermal distribution functions of the left and right thermal baths and ${\cal{T}}(\omega,T_L,T_R)$ is a generalized transmission function, possibly including interaction effects. 
For small temperature differences, (\ref{kappaQ}) reduces to $J_Q\leq\kappa_Q(T_L-T_R)$, where
$\kappa_Q \equiv \pi k^2_BT/6\hbar$ is the quantum of thermal conductance~\cite{Rego98}.

Equation (\ref{kappaQ}) is a \emph{weak} quantum bound, reflecting only unitarity,  as can be seen by inserting ${\cal{T}}(\omega>0,T_L,T_R)\leq 1$ (for bosons~\cite{noteb}) into (\ref{Landauer}).  This means that (\ref{kappaQ}) will only be saturated by systems exhibiting ballistic transport~\cite{Schwab00,Meschke06,Jezouin13}, for which ${\cal{T}}(\omega>0)=1$~\cite{Rego98}.  In nanoscale systems showing deviations from ballistic transport, as happens when the thermal energy $k_B(T_L+T_R)/2$ is greater than the energy scale(s) $\hbar\oc$ of the low-energy degrees of freedom of the subsystem hamiltonian (but much smaller than the energy scale $\hbar\omega_c$ at which the subsystem behaves classically) or when interactions are strong, (\ref{kappaQ}) will not provide a close approximation to the actual heat current.

We emphasize that (\ref{Landauer}) is not limited to weak interactions in the subsystem separating the two baths. As shown by Meir and Wingreen for charge conduction~\cite{Meir92}, and subsequently extended by others to heat transport~\cite{Wang06,Saito08,Ojanen08,Velizhanin10,Saito13}, (\ref{Landauer}) is an \emph{exact} expression for the current whenever the coupling at the left ``lead'' is proportional to that at the right [c.f. (\ref{JL})].  The identification of ${\cal{T}}(\omega,T_L,T_R)$ with nonequilibrium ($T_L\neq T_R$) correlation functions for the subsystem operators~\cite{Meir92,Wang06,Saito08,Ojanen08,Velizhanin10,Saito13}  allows us to relate the heat current to exact spectral sum rules for these correlators.

In this Letter, we derive rigorous quantum mechanical bounds on heat currents described by (\ref{Landauer}) that provide a much stronger bound than (\ref{kappaQ}) when $\hbar\oc/k_B \lesssim T_L,T_R\ll \hbar\omega_c/k_B$ or interactions are strong [this excludes the ideal gas for which (\ref{kappaQ}) is saturated since it is devoid of any energy scale]. Unlike (\ref{kappaQ}), which furnishes a bound in terms of only temperature and fundamental constants, our bounds involve a thermodynamic quantity--the $f$-sum rule---that is system-dependent and must be evaluated numerically. As with e.g., a conjectured bound on the shear viscosity of quantum fluids involving the (system-dependent) entropy~\cite{Kovtun05,Schaefer09}, it is precisely this feature that makes the bound universal, since it is saturated by \emph{all} systems with finite characteristic low-energy scales at high temperatures. In contrast, (\ref{kappaQ}) is only saturated for an ideal gas. Our bounds give physical insight into the behaviour of heat transport in a wide array of systems while replacing the difficult task of calculating time-dependent correlation functions needed for ${\cal{T}}(\omega,T_L,T_R)$ with the much easier task of calculating a simple operator expectation value (the sum rule).

{\textit{Heat current through a nanojunction coupled linearly to two boson heat baths}}---We focus here on boson (e.g., phonon and photon) heat transport  through the system shown schematically in Fig.~\ref{fig1}. The straightforward extension to electron heat transfer will be considered in another publication~\cite{Taylor15}.  We model this system by the hamiltonian
\beq \hat{H} = \!\!\! \sum_{\nu = L,R}\sum_k \left[\hbar \omega_{\nu k}\hat{b}^{\dagger}_{\nu k}\hat{b}_{\nu k} + \sum_{\vec{n}}g^{\vec{n}}_{\nu k }\hat{A}_{\vec{n}}(\hat{b}^{\dagger}_{\nu k} + \hat{b}_{\nu k})\right] + \Hss.\label{H}\eeq
Here $\hat{b}_{\nu k}$ annihilates a bath particle of energy $\hbar\omega_{\nu k}$ with quantum number (usually a momentum) $k$ in the $\nu = L/R$ bath. Their occupancies obey the equilibrium Bose distributions $n_{\nu}(\omega_{\nu k}) = [\exp(\hbar\omega_{\nu k} \beta_{\nu})-1]^{-1}$, $\beta_{\nu}\equiv (k_B T_{\nu})^{-1}$. The subsystem connecting the two baths is described by the hamiltonian $\Hss(\{\hat{A}_{\vec{n}}\})$ where $\{\hat{A}_{\vec{n}}\}$ denotes the set of operators describing the quantum states of the subsystem; $\vec{n}$ is in general a composite index denoting e.g., the quantum levels for a harmonic oscillator junction, the spin index in a spin-boson model, or pairs of  levels in a more general anharmonic junction (in which case one could choose e.g. $\hat{A}_{\vec{n}} = |i\rangle \langle j|$, $\vec{n} = \{i,j\}$ describing the transition between two levels). 
$g^{\vec{n}}_{\nu k}$ is the coupling between $\hat{A}_{\vec{n}}$ and the bath degrees of freedom.  It is characterized by the generalized spectral function~\cite{Leggett87} (not to be confused with the heat current $J_Q$ to which we always attach the subscript $Q$)
\beq J^{\vec{n},\vec{n}'}_{\nu}(\omega) \equiv \frac{4}{\hbar^2}\sum_k g^{\vec{n}}_{\nu k}g^{\vec{n}'}_{\nu k}\delta(\omega-\omega_{\nu k}).\label{I}\eeq

As shown by several authors using Keldysh methods~\cite{Saito08,Ojanen08,Velizhanin10,Saito13}, analogous to the famous Meir-Wingreen expression for the charge current through a junction~\cite{Meir92}, one can derive an \emph{exact} expression for the heat current corresponding to the model (\ref{H}), irrespective of the details of $\Hss$. When the spectral functions on the left and right sides are proportional, $J^{\vec{n},\vec{n}'}_{\nu}(\omega) \equiv 2\alpha_{\nu}\tilde{J}_{\vec{n},\vec{n}'}(\omega)$, the \emph{steady-state} heat current between the left and right baths has the particularly simple Landauer-type form [c.f. (\ref{Landauer})]
\beq J_Q \!=\! \frac{\hbar^2 \alpha\gamma}{4}\sum_{\vec{n},\vec{n}'}\!\!\;\int^{\infty}_0\!\!\!\!d\omega \omega \mathrm{Im}\chi_{\vec{n},\vec{n}'}(\omega)\tilde{J}_{\vec{n},\vec{n}'}\!(\omega)[n_L\!(\omega)-n_R(\omega)].\label{JL}\eeq
$\alpha_{\nu}$ are the dimensionless bath-junction couplings, possibly different at the left and right baths, $\alpha\equiv \alpha_L+\alpha_R$, and $\gamma\equiv 4\alpha_L\alpha_R/\alpha^2$. $\chi_{\vec{n},\vec{n'}}(\omega)\equiv \int ds \exp(i\omega s)\chi_{\vec{n},\vec{n}'}(s)$, $s\equiv t-t'$,  is the Fourier transform of the retarded correlation function for the operators $\hat{A}_{\vec{n}}$: 
\beq 
\chi_{\vec{n},\vec{n}'}(t-t')\equiv \frac{i}{\hbar}\Theta(t-t')\langle[\hat{A}_{\vec{n}}(t),\hat{A}^{\dagger}_{\vec{n}'}(t')]\rangle.
\label{chiAB} \eeq

 As we show in the Supplemental Materials, in steady-state, where the density matrix $\hat{\rho}(t)=\hat{\rho}$ is independent of time, $\chi_{\vec{n},\vec{n}'}$ obeys the well-known ``$f$-sum rule''~\cite{NozieresPines1}
\beq m_{\vec{n},\vec{n}'} \equiv  \frac{1}{\pi}\int^{\infty}_0d\omega \omega \mathrm{Im}\chi_{\vec{n},\vec{n}'}(\omega) = \frac{1}{2\hbar^2 }\langle[\hat{A}^{\dagger}_{\vec{n}'},[\hat{H},\hat{A}_{\vec{n}}]]\rangle, \label{fsum}\eeq
even for the nonequilibrium situation where the expectation value $\langle\cdots\rangle\equiv \mathrm{Tr}[\hat{\rho}(\cdots)]$ in (\ref{chiAB}) and  (\ref{fsum}) is not the thermal equilibrium one. 

{\textit{Quantum bound on the heat current}}---We now make use of the $f$-sum rule to derive a rigorous bound on the heat current (\ref{JL}). For simplicity, we will make the (natural) assumption that the form of the dissipation is the same for all of subsystem states, $\tilde{J}_{\vec{n},\vec{n}'}(\omega) = h_{\vec{n},\vec{n}'}\bar{J}(\omega)$, where $h_{\vec{n},\vec{n}'}$ is a dimensionless number that acts as a ``flag'' to enumerate the appropriate form for the operator(s) $\hat{A}_{\vec{n}}, \hat{A}_{\vec{n}'}$ coupling the thermal baths. We also restrict ourselves to the usual situation where the dissipation is ``ohmic'' with an exponential cutoff~\cite{Leggett87}:
\beq \tilde{J}_{\vec{n},\vec{n}'}(\omega) = h_{\vec{n},\vec{n}'}\omega\exp(-\omega/\omega_c).\label{ohmic}\eeq
In the Supplemental Materials we show how to generalize these bounds for other forms of dissipation.

Using (\ref{ohmic}) as well as the inequality $0\leq \hbar \omega [n_L(\omega)-n_R(\omega)]/k_B(T_L-T_R)\leq 1$ for $\omega>0$ and $T_L>T_R$, the positivity of the change in entropy, requiring $\mathrm{Im}\chi_{\vec{n},\vec{n}'}(\omega>0)\geq 0\; \forall \omega$~\cite{posdefnote},  immediately leads to the bounds
\beq J_Q \leq \frac{\pi \alpha\gamma}{4}\hbar k_B (T_L-T_R) \sum_{\vec{n},\vec{n}'}h_{\vec{n},\vec{n}'}m_{\vec{n},\vec{n}'} \label{JLineq1}\eeq
and hence, 
\beq \kappa \equiv \left.\frac{\partial J_Q}{\partial T_L}\right|_{T_L=T_R} \leq\frac{\pi \alpha\gamma}{4}\hbar k_B\sum_{\vec{n},\vec{n}'}h_{\vec{n},\vec{n}'}\left. m_{\vec{n},\vec{n}'} \right|_{T_L=T_R}.\label{kappaineq}\eeq

Equations~(\ref{JLineq1}) and (\ref{kappaineq}) are the central results of this Letter, giving rigorous bounds on the heat current through a nanojunction. Taking $h_{\vec{n},\vec{n}'}=\delta_{\vec{n},\vec{n}'}$, they admit a simple physical interpretation: $2\hbar^2 m_{A,A} = \langle [\hat{A}^{\dagger}, [\hat{H},\hat{A}]]\rangle$ is the minimum energy $E_{\mathrm{min}}$ that can be deposited in the system by the operator $\hat{A}$~\cite{Wagner66}. Our bounds can thus be understood as expressing the Heisenberg-limited time $\tau \geq \hbar/E_{\mathrm{min}}$ for quasiparticles to transport heat in $J_Q \sim k_B(T_L-T_R)/\tau$.  This provides a very natural extension of the quantum of conductance bound which instead effectively assumes $E_{\mathrm{min}} \sim k_B(T_L+T_R)$. The assumption of Ohmic dissipation is not crucial and any form of dissipation would result in sum-rule bounds of the form given above (see Supplementary Materials). In contrast, the assumption that the left and right spectral functions have the same form seems to be crucial: without this, the Meir--Wingreen expression for the current no longer involves an integral over a sign-definite function, and there is no bound in general.   

For completeness, we give the general (although not very transparent) result for the multi-level model
\beq  \Hss = \sum_nE_n|n\rangle\langle n|;\;\;\hat{A}_{ij}=|i\rangle\langle j|.\label{Hgen}\eeq
For this model, the $f$-sum rule is  
\begin{align} &m_{ij,kl}= \frac{1}{2\hbar^2}\Big\{(E_k-E_l)\left[A_{jl}\delta_{i,k}-A_{ki}\delta_{l,j}\right]\nonumber\\&+\sum_{\nu p}\Big\langle (\hat{b}^{\dagger}_{\nu p}+\hat{b}_{\nu p})\Big[\hat{A}_{jl}g^{ik}_{\nu p} + \hat{A}_{ki}g^{lj}_{\nu p}\nonumber\\ 
&-\sum_n \hat{A}_{ni}g^{nk}_{\nu p}\delta_{l,j}-\sum_n \hat{A}_{jn}g^{ln}_{\nu p}\delta_{i,k}\Big]\Big\rangle\Big\},\label{fsumrulegen}\end{align}
where $A_{ij}\equiv \langle \hat{A}_{ij}\rangle$. 
As in the spin boson and harmonic junction models we consider below, it is usually the case that the coupling $g^{\vec{n}}_{\nu k}$ assumes a simple form and one can always find a basis where 
 the relevant sum rule is greatly simplified, independent of the bath degrees of freedom. First, we briefly discuss the regimes where our sum rule bounds are saturated.

As mentioned earlier, a useful feature of our bounds is that they are expected to be saturated at high temperatures. Specifically, one needs to be at sufficiently large temperatures (but smaller than $\omega_c$) that $n_L(\omega)-n_R(\omega)\simeq k_B(T_L-T_R)/\hbar \omega$ over the range of frequencies where Im$\chi_{\vec{n},\vec{n}'}(\omega)$ is appreciable. This is the case as long as \emph{both} temperatures $T_L,T_R\gg \hbar\oc/k_B$ are larger than the scale(s) $\hbar\oc$ that characterizes the central subsystem. For a single-level harmonic junction, $\oc=\omega_0$ is just the harmonic oscillator frequency, while in the nonequilibrium spin-boson model, $\oc \sim \Delta_r$, the renormalized tunnelling frequency~\cite{Leggett87} (related, for ohmic dissipation, to the Kondo temperature $T_K$~\cite{Saito13}).

{\textit{Harmonic junction---}} We now evaluate the sum-rule bound for the case of a single-level harmonic junction
\beq \Hss = \hbar\omega_0(\hat{a}^{\dagger}\hat{a} + 1/2);\;\;\hat{A} = \hat{a}+\hat{a}^{\dagger}.\label{HO}\eeq
This model is exactly solvable and so, sum rule bounds on the current are not needed. However, it is known to saturate the ``ballistic'' bound (\ref{kappaQ}) at low temperatures, and so provides a useful system to compare this result with our new bound (\ref{JLineq1}). 

Reflecting the fact that this model admits an exact solution~\cite{Segal03,Segal05,Dhar06,Ojanen08,note},
\beq \chi(\omega) = \frac{2\omega_0/\hbar}{\omega^2-\omega^2_0-i\pi \omega_0[J_L(\omega)+J_R(\omega)]/2},\label{chi}\eeq
the $f$-sum rule (\ref{fsum}) assumes a simple form, equal to a constant: $m_{0,0} = \omega_0/\hbar$. Therefore,
\beq J_Q \leq \frac{\pi \alpha\gamma}{4} \omega_0k_B(T_L-T_R)\label{harmonicbound}.\eeq

\begin{center}
\begin{figure}
\includegraphics[width=0.98\linewidth]{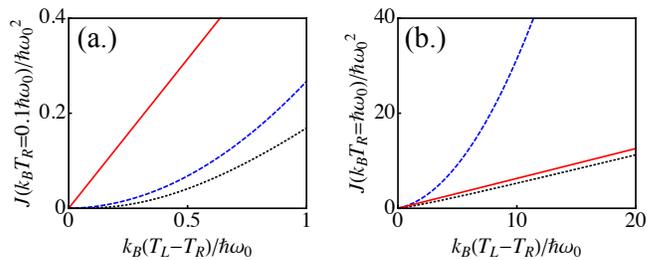}
\caption{Heat current through a harmonic junction at low (a.) and high (b.) temperatures as a function of the temperature difference $T_L-T_R$ between the left and right baths for $\alpha_L=\alpha_R=0.4$  ($\alpha=0.8,\gamma=1$). The dotted line is the exact result, the dashed blue line is the ballistic bound (\ref{kappaQ}), while the solid red line is the sum-rule bound (\ref{harmonicbound}).}\label{harmonicfig}
\end{figure}
\end{center}

\begin{center}
\begin{figure}
\includegraphics[width=0.6\linewidth]{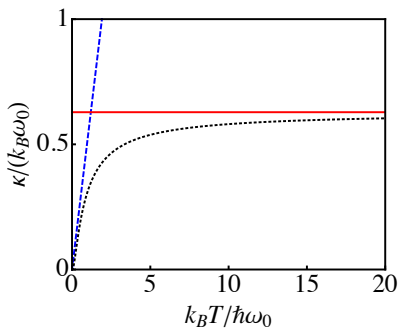}
\caption{Temperature dependence of the exact thermal conductance $\kappa$ (dotted line), the quantum of thermal conductance $\kappa_Q$ (blue dashed line), and the quantum sum-rule bound $\pi \alpha\gamma \omega_0 k_B/4$ (red solid line) for $\alpha=0.8,\gamma=1$. At low temperatures, the thermal conductance saturates $\kappa_Q$, while at high temperatures it approaches the sum-rule bound.}\label{harmonickappafig}
\end{figure}
\end{center}

Using (\ref{chi}) in (\ref{JL}) with ohmic dissipation, the heat current is plotted in Fig.~\ref{harmonicfig} and compared with the ballistic bound (\ref{kappaQ}) as well as (\ref{harmonicbound}). At low temperatures, the exact heat current nearly saturates the ballistic bound. In contrast, at higher temperatures, $T_L,T_R\gtrsim \hbar\omega_0/k_B$, the heat current is far below the ballistic bound but very nearly saturates the sum-rule bound (\ref{harmonicbound}).  This trend is more clearly seen in the thermal conductance, which we plot in Fig.~\ref{harmonickappafig} together with the thermal conductance bound (\ref{kappaineq}) and the quantum of thermal conductance.

These results confirm the picture laid out earlier: at temperatures greater than the characteristic energy scale $\hbar\omega_0$ of the subsystem, the heat current is bounded by $\omega_0$ and not the temperature. Hence, the sum-rule bounds  (\ref{JLineq1}), (\ref{kappaineq}) do far better at high temperatures than the ballistic bound (\ref{kappaQ}).  Interestingly, because the operator $\hat{A}$ in (\ref{HO}) that couples to the baths is proportional to the displacement operator $\hat{x}$ of the harmonic oscillator, adding anharmonic terms such as $V(\hat{x}) \propto \hat{x}^4$ to the susbsystem hamiltonian will not alter (\ref{harmonicbound}). For a minimal model of an anharmonic oscillator that gives rise to a nontrivial bound, we need to turn to the spin-boson model. 

{\textit{Non-equilibrium spin boson model---}}We now apply our bounds to the spin-boson model, which does not admit an exact solution (for a review of various approximation methods, see Refs.~\cite{Segal14,Boudjada14,Wang14}). For this model, 
\beq \Hss = \frac{\hbar\Delta}{2}\hat{\sigma}_x;\;\;\hat{A} = \hat{\sigma}_z,\eeq
$h_{z,z} = 1$, the $f$-sum rule is $m_{z,z}= -(\Delta/\hbar)\langle\hat{\sigma}_x\rangle$ and hence,
\beq J_Q  \leq \frac{\pi \alpha\gamma}{4} k_B(T_L-T_R) \left(- \Delta\langle\sx\rangle \right)\label{JLineq2}\eeq
and
\beq  \kappa\leq \frac{\pi \alpha\gamma}{4} k_B\left.\left(-\Delta\langle\sx\rangle \right)\right|_{T_L=T_R}.\label{kappaineq2}\eeq
The above results are most easily obtained from $m_{z,z}=\langle[\hat{\sigma}_z,[\hat{H},\hat{\sigma}_z]]\rangle/2\hbar^2$; they also result from the more general (\ref{Hgen}) and (\ref{fsumrulegen}) by setting $g^{ij}_{\nu p} = g_{\nu p}[\delta_{i,1}\delta_{j,2} + \delta_{i,2}\delta_{j,1}]$ (corresponding to coupling between the baths and the $\hat{\sigma}_x$ operator),  $\hbar\Delta = E_2-E_1$, and $m_{x,x} = (m_{12,12}+m_{12,21}+m_{21,12}+m_{21,21})$ (which follows from $\hat{\sigma}_x = |1\rangle\langle2|+|2\rangle\langle 1|$), and afterwards rotating the spin basis: $\hat{\sigma}_z \to \hat{\sigma}_x$, $\hat{\sigma}_x\to -\hat{\sigma}_z$. The dependence on the bath degrees of freedom that arises in (\ref{fsumrulegen}) cancels out in $m_{z,z}$. 

\begin{center}
\begin{figure}
\includegraphics[width=0.64\linewidth]{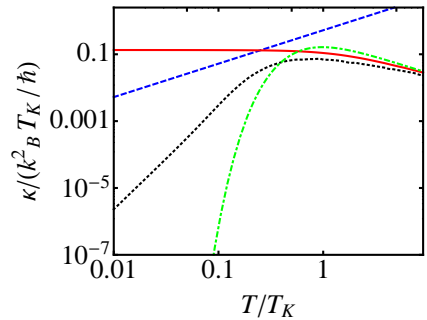}
\caption{Temperature dependence of the thermal conductance bounds (red solid and blue dashed lines) for $\alpha=0.2, \gamma=1$. The QMC value for $\kappa$~\cite{Saito13} (dotted line) saturates the sum-rule bound (red solid line) for temperatures greater than the Kondo temperature $T_K$, but is always much less than the quantum of thermal conductance (blue dashed line). The Bloch--Redfield value (green dot-dashed line) violates the bound at intermediate temperatures. The sum-rule bound (\ref{kappaineq2}) is calculated using QUAPI for $\omega_c=10\Delta$ and the numerical algorithm developed in Ref.~\cite{QUAPIGFP}.}\label{NESBfig}
\end{figure}
\end{center}

In contrast to the exactly solvable harmonic junction model, the right-hand sides of these bounds must be evaluated numerically. This is straightforward to do exactly using the quasi-adiabatic path-integral (QUAPI) approach~\cite{QUAPIMM,QUAPIGFP} and in Fig.~\ref{NESBfig}, we plot the sum-rule conductance bound (\ref{kappaineq2}) for $\alpha=0.2,\gamma=1$ together with the quantum of thermal conductance $\kappa_Q$. For comparison we also show the thermal conductance taken from a recent quantum Monte Carlo (QMC) calculation~\cite{Saito13} as well as the weak-coupling Bloch--Redfield (BR) result which, in the limit $\omega_c\gg \Delta$ is given by~\cite{Boudjada14} $\kappa = (\hbar^2\pi/8k_B)(\Delta^3/T^2)\alpha\gamma/\sinh(\beta\hbar \Delta)$.  

As expected, the sum-rule bound is saturated from below by the (in principle) exact QMC result at temperatures above the Kondo temperature~\cite{Leggett87,Saito13} $T_K\equiv g_{\alpha}(\hbar \Delta/k_B) (\Delta/\omega_c)^{\alpha/(1-\alpha)}$ for $\alpha\leq 1$ and 0 for $\alpha>1$. $g_{\alpha} \equiv [\Gamma(1-2\alpha)\cos(\pi\alpha)]^{1/2(1-\alpha)}$, with $\Gamma$ the Gamma function. For $T\gg T_K$, both $\kappa$~\cite{Saito13} and $\langle\hat{\sigma}_x\rangle$~\cite{Weissbook} scale as $(T/T_K)^{2\alpha-1}$, as required by our sum rule bound.  The BR thermal conductance violates the sum rule bound at intermediate temperatures $T\sim T_K$. For higher temperatures, it scales as $1/T$ and lies below the sum-rule bound as well as the QMC result. At weaker coupling, $\alpha\lesssim 0.1$, the BR value is below the bound for all temperatures.

The sum rule bounds (\ref{JLineq2}) and (\ref{kappaineq2}) show the crucial role played by the \emph{off-diagonal} coherence $\langle\sx\rangle$ in the heat current. For small coupling $\alpha$, $\langle\hat{\sigma}_x\rangle$ is independent of $\alpha$ and the heat current will increase linearly with the coupling. Ultimately, however, for values of this coupling beyond the so-called Toulouse point $\alpha=0.5$~\cite{Leggett87,LeHur08}, the spin subsystem is strongly entangled with the environment, the coherence $\langle\hat{\sigma}_x\rangle$ is greatly suppressed, and the thermal conductance becomes asymptotically small~\cite{Saito13}.

 {\textit{Conclusions---}}We have derived rigorous quantum bounds on heat transport through nanojunctions that extend the well-known quantum of thermal conductance bound to better describe interacting systems.   These bounds are universal in the sense that for any system with characteristic low-energy energy scales, the actual heat current saturates our bounds for temperatures above these scales. In this way, our bounds will always outperform the ballistic bound at these temperatures, while at temperatures below these scales, the ballistic bound is generally stronger [see Figs.~\ref{harmonickappafig} and \ref{NESBfig}]. Ideal gases that saturate the ballistic bound are devoid of such scales and our bound will do worse (it is formally divergent!). For all other systems with energy scales introduced by e.g. interactions or an ultraviolet cutoff (as in few-level systems), our bounds are well-saturated at intermediate temperatures, opening the door to investigating room-temperature heat transport in nanoscale devices when interactions are strong. 
It will be particularly interesting to revisit foundational questions of entropy flow~\cite{Pendry83,Blencowe00} when interactions are strong and also to study how interactions can be used to engineer quantum heat devices such as thermal diodes~\cite{Segal05b}. 
 
 {\textit{Acknowledgments}}---We thank Keiji Saito and Takeo Kato for providing us with the QMC data from Ref.~\cite{Saito13} used in Fig.~\ref{NESBfig} and Andrei Golosov  for providing us with his QUAPI routine described in Ref.~\cite{QUAPIGFP}.  Support from the Canada Research Chair Program and an NSERC discovery grant is acknowledged.


\clearpage

\section{Supplemental Material}
{\textit{Spectral properties of the nonequilibrium correlation function---}}
Here we discuss the spectral properties of the in-general nonequilibrium retarded correlation function for the operators $\hat{A}$ and $\hat{B}$ [note that for brevity, we adopt a slightly different subscript notation in Eq.~(6) in the main text]:  
\beq \chi_{A,B}(t,t')\equiv \frac{i}{\hbar}\Theta(t-t')\langle[\hat{A}(t),\hat{B}^{\dagger}(t')]\rangle.\label{chiABs}\eeq
The expectation value that enters this expression can be expressed in terms of the eigenstates $|a\rangle, |b\rangle$ of the hamiltonian $\hat{H}$ as
\begin{align} \langle \hat{Q}(t)\rangle &= \sum_{ab}\rho_{ab}\langle a|\hat{Q}(t)|b\rangle\nonumber\\
&=\sum_{ab}\rho_{ab}e^{\tfrac{it}{\hbar}(E_a-E_b)}\langle a|\hat{Q}|b\rangle. \label{evs}\end{align}
Here e.g., $E_a$ is the eigenvalue corresponding to the eigenstate $|a\rangle$: $\hat{H}|a\rangle = E_a |a\rangle$.  $\rho_{ab}$ are the matrix elements of the in general time-dependent density matrix by $\hat{\rho}=\sum_{ab}\rho_{ab} |b\rangle\langle a|$; in thermal equilibrium, it assumes the usual form $\rho_{ab} = \delta_{a,b}\exp(-\beta E_a)/{\cal{Z}}$, where ${\cal{Z}} = {\rm Tr}[\exp(-\beta\hat{H})]$ is the partition function.  
We will generically be interested in the nonequilibrium situation, where $T_L\neq T_R$.  As with the equilibrium situation, however, in steady-state, where the left-hand-side of (\ref{evs}) is time-independent, the summation in this expression collapses and one can replace $\rho_{ab}$ with $\rho_{aa}\delta_{a,b}$.

Using (\ref{evs}), one can write down the spectral representation for $\chi_{A,B}$ using the exact eigenstates and eigenvalues of the full hamiltonian.  Doing this, one obtains
\begin{align} &\chi_{A,B}(t,t')=\nonumber\\& \frac{i}{\hbar}\Theta(t-t')\sum_{abc}\rho_{ac}\Big[e^{\tfrac{it}{\hbar}(E_a-E_b)-\tfrac{it'}{\hbar}(E_c-E_b)}\langle a|\hat{A}|b\rangle \langle b|\hat{B}^{\dagger}|c\rangle\nonumber\\& -e^{\tfrac{it'}{\hbar}(E_a-E_b)-\tfrac{it}{\hbar}(E_c-E_b)}\langle a|\hat{B}^{\dagger}|b\rangle\langle b|\hat{A}|c\rangle\Big].\label{chiAB2s}\end{align}

We suppose that the system settles down to a steady-state configuration after some characteristic time $t^*$.  For $t,t'$ both much larger than this time scale, the two-time retarded correlator $\chi_{AA}(t,t')$ only depends on the difference between $t$ and $t'$:
\beq \chi_{A,B}(t,t') \to \chi_{A,B}(t-t').\eeq
This means that the summation over the indices $a,b,c$ in (\ref{chiAB2s}) collapses, with $c=a$.  Fourier transforming the resulting expression gives
\begin{align}& \chi_{A,B}(\omega) =\nonumber\\&\frac{1}{\hbar}\sum_{ab}\rho_{aa}\left[
\frac{\langle a|\hat{B}^{\dagger}|b\rangle\langle b|\hat{A}|a\rangle}{\hbar \omega + E_{ba} + i0^+}
- \frac{\langle a|\hat{A}|b\rangle\langle b|\hat{B}^{\dagger}|a\rangle}{\hbar \omega - E_{ba} + i0^+} 
\right],
\label{FourierchiAB}
\end{align}
where $E_{ba}\equiv E_b-E_a$.  The imaginary part that enters the expression for the heat current is
\beq \mathrm{Im} \chi_{A,B}(\omega) = \frac{\pi}{\hbar}\sum_{ab}(\rho_{aa}-\rho_{bb})\langle a|\hat{A}|b\rangle\langle b|\hat{B}^{\dagger}|a\rangle 
\delta(\hbar \omega -E_{ba}).
\label{ImchiAB}
\eeq

Multiplying (\ref{ImchiAB}) by $\omega$ and integrating yields
\begin{align}& \frac{1}{\pi}\int^{\infty}_{-\infty} d\omega \omega \mathrm{Im} \chi_{A,B}(\omega) =\frac{1}{\hbar^2}\sum_{ab} (\rho_{aa}-\rho_{bb})E_{ba}\times \nonumber\\
&\langle a|\hat{A}|b\rangle\langle b|\hat{B}^{\dagger}|a\rangle=\frac{1}{\hbar^2}\sum_{a} \rho_{aa}\langle a|[\hat{B}^{\dagger},[\hat{H},\hat{A}]]|a\rangle.  \label{fsums}\end{align}
Making use of the fact that $\mathrm{Im}\chi_{A,B}(-\omega) = -\mathrm{Im}\chi_{A,B}(\omega)$, setting $\hat{A}=\hat{A}_{\vec{n}}, \hat{B} = \hat{A}_{\vec{n}'}$, and recalling that $\rho_{ab}$ can be replaced by $\rho_{aa}\delta_{a,b}$ in (\ref{evs}) in steady-state gives the $f$-sum rule shown by Eq.~(7) in the main text.  

{\textit{Other forms of dissipation---}}For simplicity, we assumed ohmic dissipation in the main text.  Here we show how to extend our results to other forms of dissipation.   

Any form of dissipation $\tilde{J}_{\vec{n},\vec{n}'}(\omega)$ leads to a bound on the heat current: one only needs to find a function $F_{\vec{n},\vec{n}'}(\omega)\geq \tilde{J}_{\vec{n},\vec{n}'}(\omega)$ $\forall \omega$ such that $\int^{\infty}_0 d\omega F_{\vec{n},\vec{n}'}(\omega)\mathrm{Im}\chi_{\vec{n},\vec{n}'}(\omega)$ can be calculated (this can always be done).  For $F(\omega) = \omega^l$, where $l$ is an integer and we have dropped the $\vec{n},\vec{n}'$ subscript for clarity, such sum rules are straightforwardly given as a commutator expression, as in (\ref{fsums}).  In all cases of current interest, including ``super-ohmic'' dissipation with $l>1$, Lorentzian, and Gaussian forms, the spectral function can be bounded by such a simple polynomial expression.  At the same time, for  e.g., a Lorentzian spectral functions of the form $\tilde{J}(\omega) = \omega^l[1+\eta(\omega-\omega_R)^2]^{-1}\leq \omega^l$, improved bounds can be obtained by directly evaluating $\int^{\infty}_0 d\omega \tilde{J}(\omega)\mathrm{Im}\chi_{\vec{n},\vec{n}'}(\omega)$.  This can be done by applying the operator-product expansion of Kadanoff, Wilson, and Polyakov~\cite{OPEs}.  (See Refs.~\cite{Poggio76s,Braaten10s} for examples of such ``smeared'' sum rule calculations.)  The same technique can be used to calculate sum rules for non-analytic $F(\omega)$, including power-law forms with non-integer exponents.

For these more general forms of dissipation, in addition to the high-temperature criterion outlined in the main text, in order for the exact heat current to saturate the resulting sum rule bound, one needs the spectral function to be approximately power-law over the range of frequencies where $\mathrm{Im}\chi_{\vec{n},\vec{n}'}$ is appreciable.  For the Lorentzian form discussed above, this means that $\mathrm{Im}\chi_{\vec{n},\vec{n}'}$ must be peaked in the vicinity of the resonance frequency $\omega_R$.  Even if this condition is not satisfied, we emphasize that a.) the bound remains rigorous and b.) one can derive improved sum rule bounds along the lines suggested above, and so at sufficiently large temperatures, $\hbar \oc\ll k_BT_L,k_BT_R\ll \hbar\omega_c$, the appropriate sum-rule bound will always be very nearly saturated.


\begin{thebibliography}{99}
\bibitem{noteb} For bosons, ${\cal{T}}(\omega) \propto \Theta(\omega)$ is zero for $\omega<0$. For fermions, this constraint is absent.
\bibitem{note} Note that Ref.~\cite{Ojanen08} obtains a more general expression than the one given here, allowing for a complex admittance $Y_{\nu}(\omega)$, the real part of which is related to the spectral function by (using their notation) Re$Y_{\nu}(\omega>0) = (\pi\hbar/4\omega)J_{\nu}(\omega)/(MI_0)^2$.
\end{thebibliography}
\end{document}